# Nuclear Physics Programs for the Future Rare Isotope Beams Accelerator Facility in Korea

MOON Chang-Bum*

Hoseo University, Asan, Chung-Nam 336-795, Korea

30 August, 2015

**Abstract**

We present nuclear physics programs based on the planned experiments using rare isotope beams (RIBs) for the future Korean Rare Isotope Beams Accelerator facility; RAON. This ambitious facility has both an Isotope Separation On Line (ISOL) and fragmentation capability for producing RIBs and accelerating beams of wide range mass of nuclides with energies of a few to hundreds MeV per nucleon. Low energy RIBs at $E_{lab}$ = 5 to 20 MeV per nucleon are for the study of nuclear structure and nuclear astrophysics toward and beyond the drip lines while higher energy RIBs produced by in-flight fragmentation with the re-accelerated ions from the ISOL enable to explore the neutron drip lines in intermediate mass regions.

The planned programs have goals for investigating nuclear structures of the exotic nuclei toward and beyond the nucleon drip lines by addressing the following issues: how the shell structure evolves in areas of extreme proton to neutron imbalance; whether the isospin symmetry maintains in isobaric mirror nuclei at and beyond the drip lines, how two-proton radioactivity affects abundances of the elements, what the role of the continuum states including resonant states above the neutron- and proton-decay threshold in exotic nuclei is in astrophysical nuclear reaction processes, and how the nuclear reaction rates triggered by unbound proton-rich nuclei make an effect on rapid proton capture processes in a very hot stellar plasma.

* cbmoon@hoseo.edu

## 1 Introduction

Atomic nuclei are complex quantum-mechanical many-body systems with a finite number of nucleons; protons, Z, and neutrons, N. As a result, shell structure of the nuclei reveals discretely according to the Z and N numbers. The question of how shell structure develops in the finite quantum many-body systems has been a common problem among various disciplines; nuclear physics, atomic physics, condensed matter physics, molecular physics, and biophysics. The shell structure of the atomic nucleus is one of the cornerstones for a comprehensive understanding of the many-body quantum mesoscopic system. The new isotope beams offer an opportunity for approaching and mapping regions of the drip lines, and helps the study of nuclear many-body quantum stability toward the proton and neutron drip lines. The study of nuclear structure including nuclear astrophysics is going through a new era owing to the development of rare isotope beams (RIBs) accelerators and a new generation of sophisticated detector systems,

World-class accelerator facilities to produce RIBs being operated and planned in the world are: SPIRAL I and II(Système de Production d'Ions Radioactifs en Ligne) at GANIL(Grand Accélérateur National d'Ions Lourds) in France [1], RIBF (Radioactive Ion Beams Factory) at RIKEN in Japan [2], NSCL(National Super Conducting Cyclotron Laboratory) at Michigan State University (MSU) in USA [3], which will become the FRIB (Facility for RIBs) at MSU [4], FAIR at GSI (Gesellschaft fur Schwerionenforschung mbH) in Germany [5], ISAC I and II (Isotope Separator and ACelerator) at TRIUMF(Tri-University Meson Facility) in Canada [6], and REX and HIE-ISOLDE at CERN [7].

The future of nuclear physics depends to a large extent on the planning of the new facilities. In this respect, the Korean Rare Isotope Beams Accelerator facility to be built (**RAON**; **R**are isotope beams **A**ccelerat**O**r **N**ational facility)* stands on the heart of the future of nuclear physics in the world. We report on nuclear physics programs at the RAON facility with a goal for exploring new areas of the nuclei and the extreme areas of astrophysical nuclear reaction processes.

## 2 An overview of the shell structure evolution

To observe the behavior related to a fundamental characteristic of nuclear structure, it is better to draw a systematics of experimental observables across the nuclear chart [8, 9, 10, 11]. Especially, illuminating are the systematics of the first $2^+$ excited states of isotopic and isotonic chains which span the major shell closure. Figure 1 describes a nuclear shell structure development; (1) upper part shows the nuclear energy level sequences based on the shell model theory and (2) lower part shows the systematics of the first excited $2^+$ states experimentally observed in even-even nuclei in the vicinity of shell closures [11]. We notice that large energy gaps are observed at $^{40}$Ca, $^{48}$Ca, $^{56}$Ni, $^{68}$Ni, and $^{90}$Zr, which are known as a doubly-magic nucleus. Dynamical structural changes along the lines of both Z and N = 20, 28, and 40 isotones and isotopes have been main subjects of the studies in nuclear structure physics.

Excluding characteristics of the doubly magic nuclei, we notice the following features revealed in the first excited $2^+$ states: First, shell gap at N = 20 is more reinforced with Z = 14 and 16 than that with Z = 18, and weakened below Z = 14 and above Z = 20. Second, the shell closure at Z = 20 (Ca isotopes) is more enhanced with N = 16 and 32. It is worthwhile to know that the shell gap at Z = 8 (O isotopes) shows a semi-doubly shell closure at N = 14 and 16, respectively. According to the level sequences as shown in Fig. 1, N = 14 and 16 occupy sequentially $1d_{5/2}$ and $2s_{1/2}$, leading to a sub-shell gap. Similarly, N = 32 and 34 form a sub-shell gap by occupying $2p_{3/2}$ and $2p_{1/2}$ orbitals, respectively. If we follow this tendency for a shell gap, we expect that $^{34}$Ca, with N = 14, and $^{54}$Ca, with N = 34, would have a semi-doubly magic character. The study of $^{34}$Ca, known as a nucleus with two proton radioactivity, is one of the subjects for our plans [11]. Third, the shell gap at N = 40 is enhanced only at Z = 28. This implies that the nuclei with N = 40 favor collective rather than spherical character for ground states. Fourth, proton number 40 (Zr isotopes) develops a semi-doubly magic character at N = 56 and at N = 58 that are also the sub-shell gap numbers.

By focusing on the harmonic oscillator shell closure numbers; 8, 20, 40, and 70, we notice the following particular responses to the sub-shell gaps: Z = 8 responds to N = 14 and 16; Z = 20 does to N = 32 and 34; and Z = 40 does to N = 56 and 58, which lead to a semi-doubly shell closure. For N = 40, we raise a question, *'why does not it respond to Z = 32 and 34 for developing a semi-doubly shell closure?'*. This problem may be intimately connected with concepts of the shape coexistence in the $^{72}$Ge (Z = 32) and $^{74}$Se (Z = 34) nuclei. In contrast, both of N = 70 and Z = 70 shell numbers have no shell gap partners leading to a semi-doubly shell closure. For instance, $^{120}$Sn (Z = 50 and N = 70) shows no evidence for a semi-double shell closure, proving that $1h_{11/2}$ orbital remains in a strong intruder state. In this regard, it is questionable that $^{110}$Z (Z = 40 and N = 70) would reveal a semi-magic character.

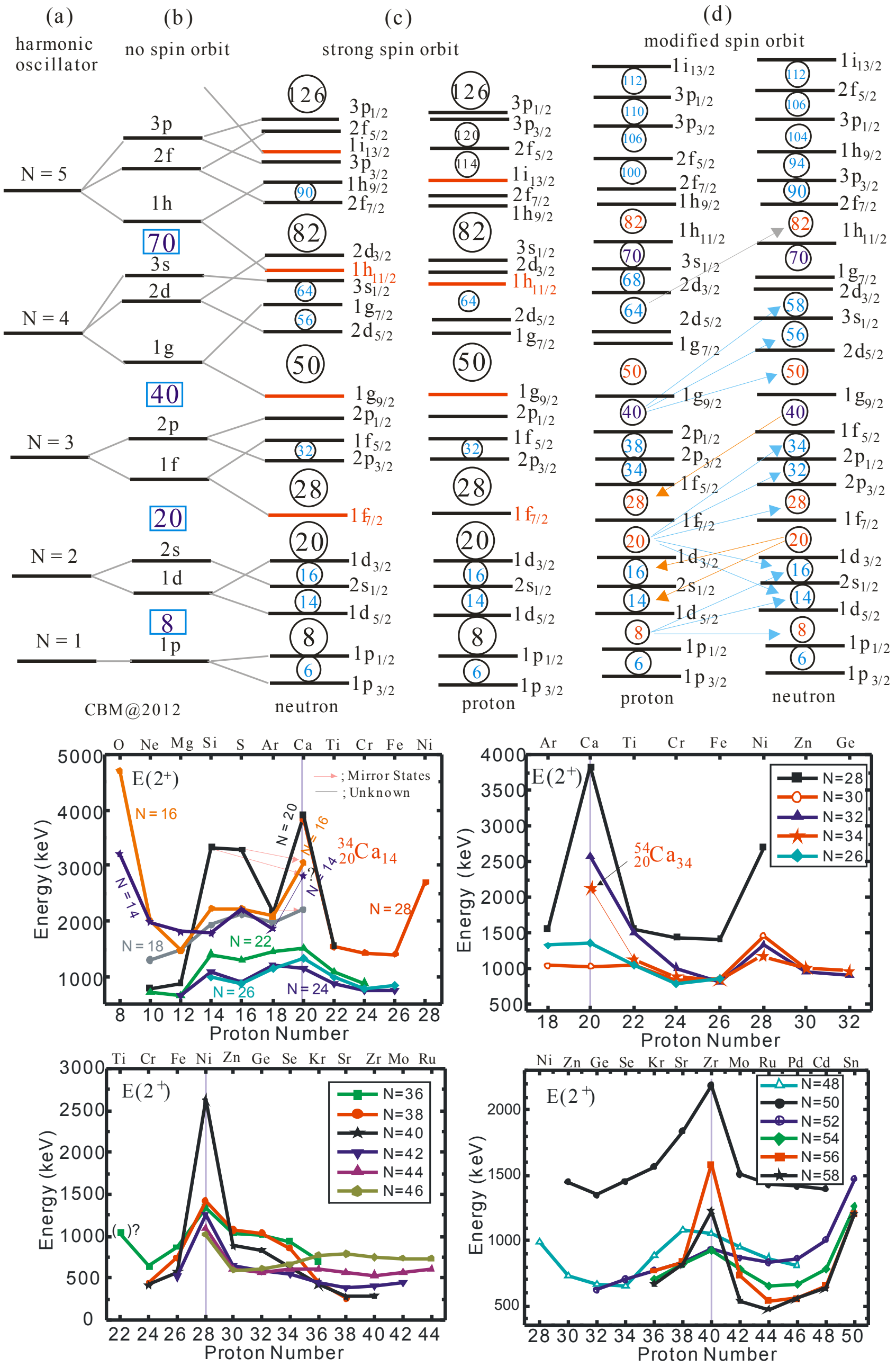

Fig. 1. (Upper panel) (a) The single particle energies of a harmonic oscillator potential as a function of the oscillator quantum number N. (b) A schematic representation of the single-particle energies of a Woods-Saxon potential. (c) A schematic illustration of the level splitting due to the spin-orbit coupling term. The numbers at the energy gaps are the subtotals of the number of particles represented by Nj = 2j + 1 of identical particles that can occupy each state. Note that the levels are numbered serially by a given orbital quantum number. (d) Modified energy levels according to the existing experimental data for neutron-rich region. The level pattern given represents only qualitative features. The arrows represent possible combinations of the proton and neutron (sub) shell gap numbers for developing a semi-double shell closure. (Below) energy level systematics for the first excited $2^+$ states in even-even nuclei as a function of proton numbers [11]. Data are taken primarily from NNDC [12].

## 3 A brief overview of nuclear experiments

The general experimental technique for nuclear physics includes mass measurements, stopped beam beta(β)-ray measurements, and nuclear reactions [13]. The mass measurements allow us to obtain basic information on nuclear structure, nuclear astrophysics, and fundamental interactions [13]. The nuclear masses are a direct reflection of the energies of the nuclei. In equilibrium, a system trends toward the lowest energy states and the transition to lower energy states releases energy, providing a source to power and to explode stars [14]. The stability of the nuclei against the various modes of radioactive decay can easily be understood in terms of the liquid drop model mass formula: $M(Z, A) = (A - Z)m_n + Z(m_p+m_e) - a_1A + a_2A^{2/3} + a_3(A/2-Z)^2/A + a_4Z^2/A^{1/3} + a_5\delta(A)$, here $a_1$, $a_2$, $a_3$, $a_4$, and $a_5$ are coefficients due to volume, surface, asymmetric, Coulomb, and pairing energy terms, respectively. The pairing term $\delta(A)$ corresponds to $1/A^{3/4}$ for odd-odd nuclei, 0 for odd-even nuclei, and $-1/A^{3/4}$ for even-even nuclei [15].

The stopped beam β-γ spectroscopy allows us to measure gamma rays following beta-decay of excited states into the daughter nuclei. Besides, the stopped beam spectroscopy with sophisticated detector systems offers information on exotic decay modes such as β-delayed proton(s) or neutron(s) emission in the nuclei toward the drip lines [16]. Data on β-delayed neutron emission plays a key role in understanding of the abundances of the elements as it affects the pathways of the rapid neutron capture processes (r-process) [17]. Both the detailed stopped beam and the in-beam spectroscopic techniques provide complementary data on the location and the ordering of single particle states for exotic nuclei of interest. They also enable us to deduce radiative neutron capture on the very neutron-rich nuclei impossible to access in direct measurements.

It is difficult and/or impossible to directly measure the thermal nuclear reaction rates based on (p, γ), (p, α), and (α, p) with RIBs such as; $^{14}O$, $^{15}O$, $^{17}F$, $^{18}Ne$, $^{22}Mg$, $^{23}Mg$, $^{26}Al$, $^{25}Si$, and $^{44}Ti$,, because secondary beams are limited to both intensity and production from ISOL. Instead, indirect measurements based on elastic resonance scatterings, transfer reactions, and Coulomb excitations provide information on the reaction rates for a given reaction system. The cross sections for the capture reactions can be determined by Coulomb dissociation based on the inverse photo-dissociation reactions [18]. The elastic scattering reactions are needed to study resonant states at higher excitation energies and provide information on the Coulomb amplitude and the nuclear amplitude [13, 19]. The inelastic scatterings offer the properties of states in a compound nucleus where decay by particle emission to an excited state is possible. Among direct nuclear reactions, the single-nucleon removal (knockout or breakup) reactions with heavy projectile ions at intermediate energies (100~300 MeV per nucleon) have become a specific and quantitative tool for studying single-particle occupancies and correlation effects in the nuclear shell model. Charge exchange reactions are another method to measure Gamow-Teller strength compared to the usual β–decay study. While the measurements of the β–decay are limited with Q values, the charge exchange reactions do not have such a limit. The charge exchange reactions will play another important role in studying the properties of exotic nuclei.

For nuclear astrophysics, the proposed experiments are mainly focused on the rapid proton capture reactions (rp-process) in a very hot stellar plasma. The rp-process has been known to be considerably complex due to the interplay of proton captures, decays, possibly photo-disintegrations, and particle induced reactions [20]. The important factors that influence the rp-process nuclear reaction networks above $Z \geq 32$ include the proton-capture reaction rates and their inverse photo-disintegration rates, and the β-decay and electron-capture rates [20]. We should remember that nuclear deformations significantly affect both the r-process and the rp-process reactions.

## 4 The characteristics of the RAON facility

To produce radioactive ion beams, there are two main approaches; ISOL production [19, 21] and In-Flight production [22]. The RAON facility has both an ISOL and fragmentation capability. The efforts to create, separate, and study radioactive nuclides are being undertaken for the coming decade. This is an ambitious project to build a multi-beam facility capable of producing and accelerating beams of a wide mass range of nuclides with energies of a few to hundreds MeV per nucleon. See Fig. 2. The KOBRA (KOrea Broad acceptance Recoil spectrometer and Apparatus) and the IFFS (In-Flight Fragmentation Separator) facilities will provide the selection and identification of the beam-like particles as well as target-like ones, and with the combination of γ-ray detectors array offer spectroscopic information on internal structures of the nuclei to be investigated.

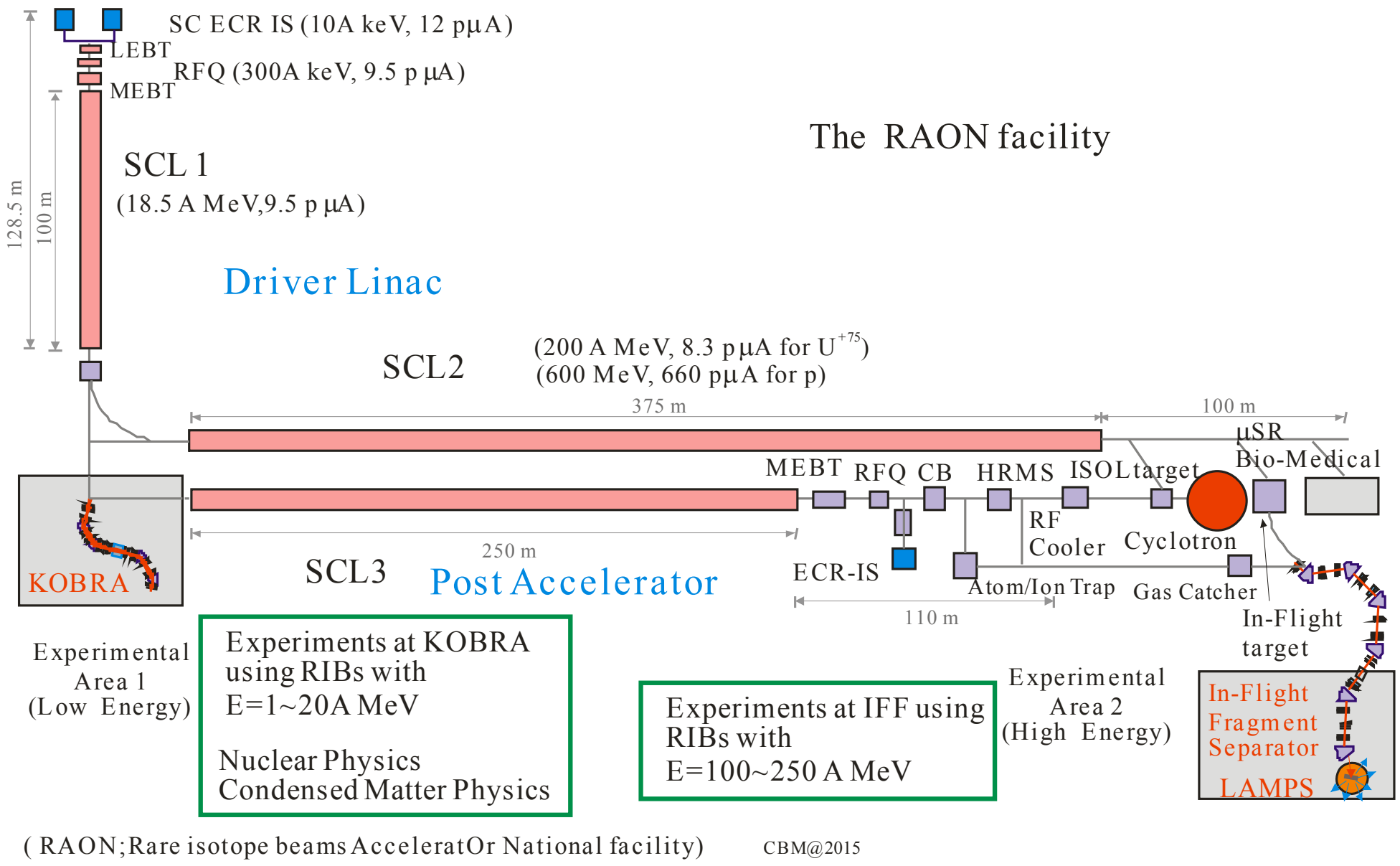

Fig. 2. Layout of the RAON facility [23].

The KOBRA facility is divided into two stages by consisting of a series of magnetic dipoles, quadrupoles, and Wien filters as shown in Fig. 3 [23]. The stage 1 is utilized to produce the low energy RIBs via multi-nucleon transfer reactions at about 20 MeV per nucleon or via direct one or two nucleon transfer reactions at a few MeV per nucleon [24]. It is worthwhile to know that it is designed to reject the primary beams for the studies of thermonuclear, namely direct capture reactions under high temperature plasma conditions in hot stars. The stage 2 placed at down-stream of the stage 1 is employed to separate the fragments following the reactions of RIBs on a reaction target, or to reject the primary beam in the same manner as for the stage 1.

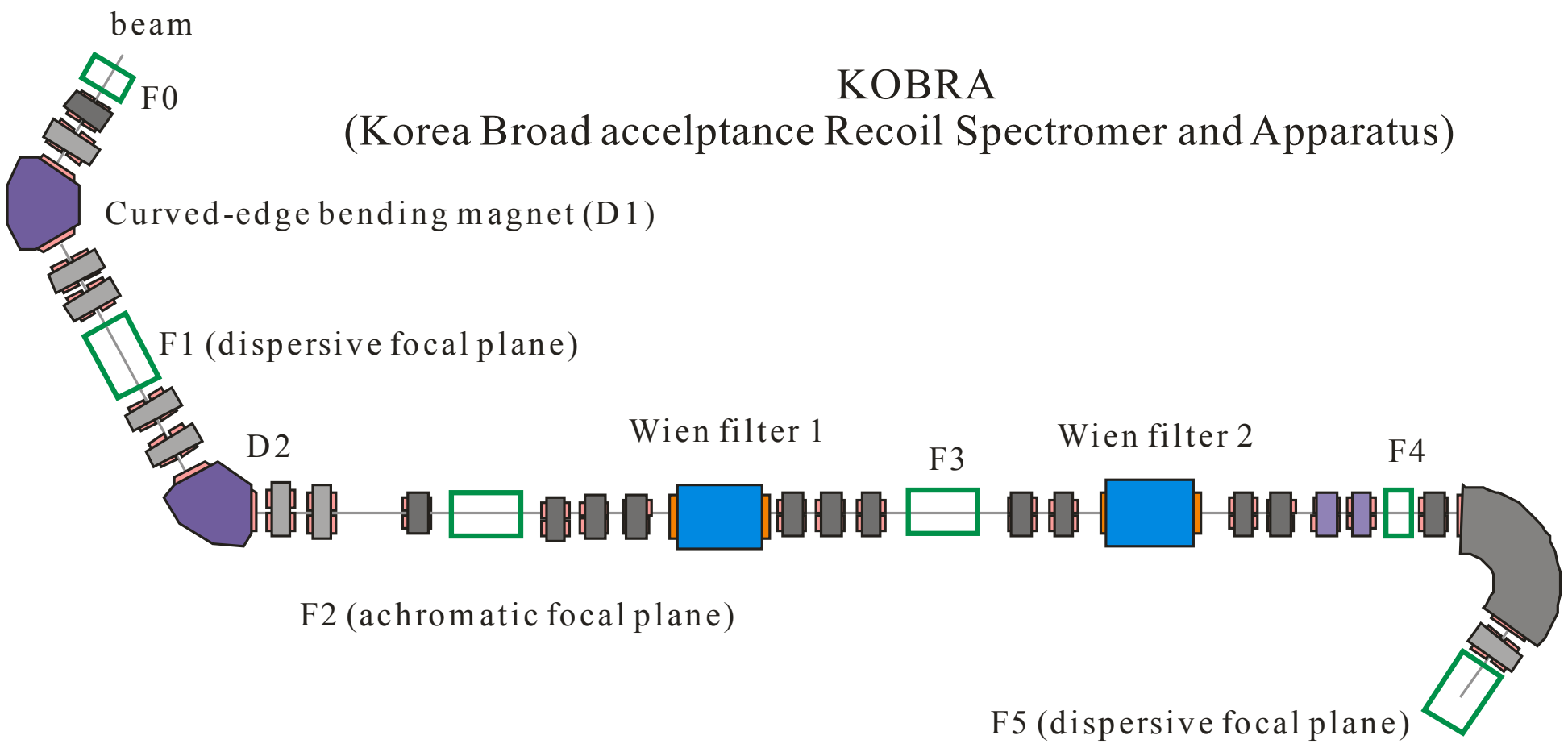

Fig. 3. Layout of KOBRA. This facility is divided into a stage 1 (F0 - F3) and a stage 2 (F3 – F5). Curved-edge shape bending magnets are utilized to minimize the high order aberration [23, 24].

At the KOBRA facility, the lower energy re-accelerated ISOL rare isotope beams with energies 5 to 15 MeV per nucleon can be separated and identified. Such low energy reaccelerated RIBs are suitable for producing the nuclear reactions such as; Coulomb excitations, nuclear fusion-evaporation reactions, elastic resonance scatterings, inelastic scatterings, one or two nucleon transfer reactions, and direct capture reactions. The higher energy RIBs with energies of 100 up to 250 MeV per nucleon will be produced by nuclear fragmentation at the IFFS. High quality and intense RIBs combined with re-accelerated ISOL beams including high efficient detector systems will provide unique experimental possibilities to study the very neutron-rich and proton-rich nuclei toward and beyond the drip lines.

The KOBRA and the IFFS have two complex detector systems: One is located at the target position for detection of reaction products including light-charged particles and γ rays; and the second one is positioned at focal plane for detection and identification of heavy recoiled nuclei as well as for spectroscopic study of delayed activity. Conjunction of detector systems with a polarized spin target (or polarized beam) is desirable for studying spin-orbital interactions.

## 5 The planned experiments

In the following we describe briefly some proposed experiments aiming at the RAON facility. More detailed descriptions are found in [11].

### A Study on two proton capture reactions.

The proton drip line imposes a constraint on the reaction path of the rp-process. As shown in Fig. 4, many proton-rich nuclides near the Z = N line are unbound to proton decay. If we assume an immediate proton-decay for these proton unbound nuclei, we expect that no further proton-capture proceeds. Thus further processing depends on the β-decay of the last proton stable isotone. This bottleneck is called a waiting point. However, if the lifetime of a proton unstable nucleus is appreciably long due to a high Coulomb barrier, 2p-capture reactions on the last proton bound isotone would be possible [20]. The 2p-capture reactions allow to bridge the single proton unstable nucleus to a proton bound nucleus. This plan aims at measuring the 2p capture reaction rates combined with the proton resonance scatterings for the associated unbound nuclei. The 2p-capture reactions to be studied are summarized in Table I.

We introduce, as an example in this field, a proposal for investigating the unbound states of $^{73}$Rb. This proposal aims to measure the unbound states of the proton radioactive $^{73}$Rb through proton elastic resonance scatterings with $^{72}$Kr on a hydrogen thick target. $^{72}$Kr is a waiting point where the proton capture is followed by the instantaneous emission of a proton by the proton unbound $^{73}$Rb nucleus. This waiting point, however, as already pointed out, can be bypassed by a two-proton radiative capture reaction if the nucleus $^{74}$Sr is bound. Such alternative paths can be estimated by the calculations using the properties of the intermediate unbound nucleus, $^{73}$Rb.

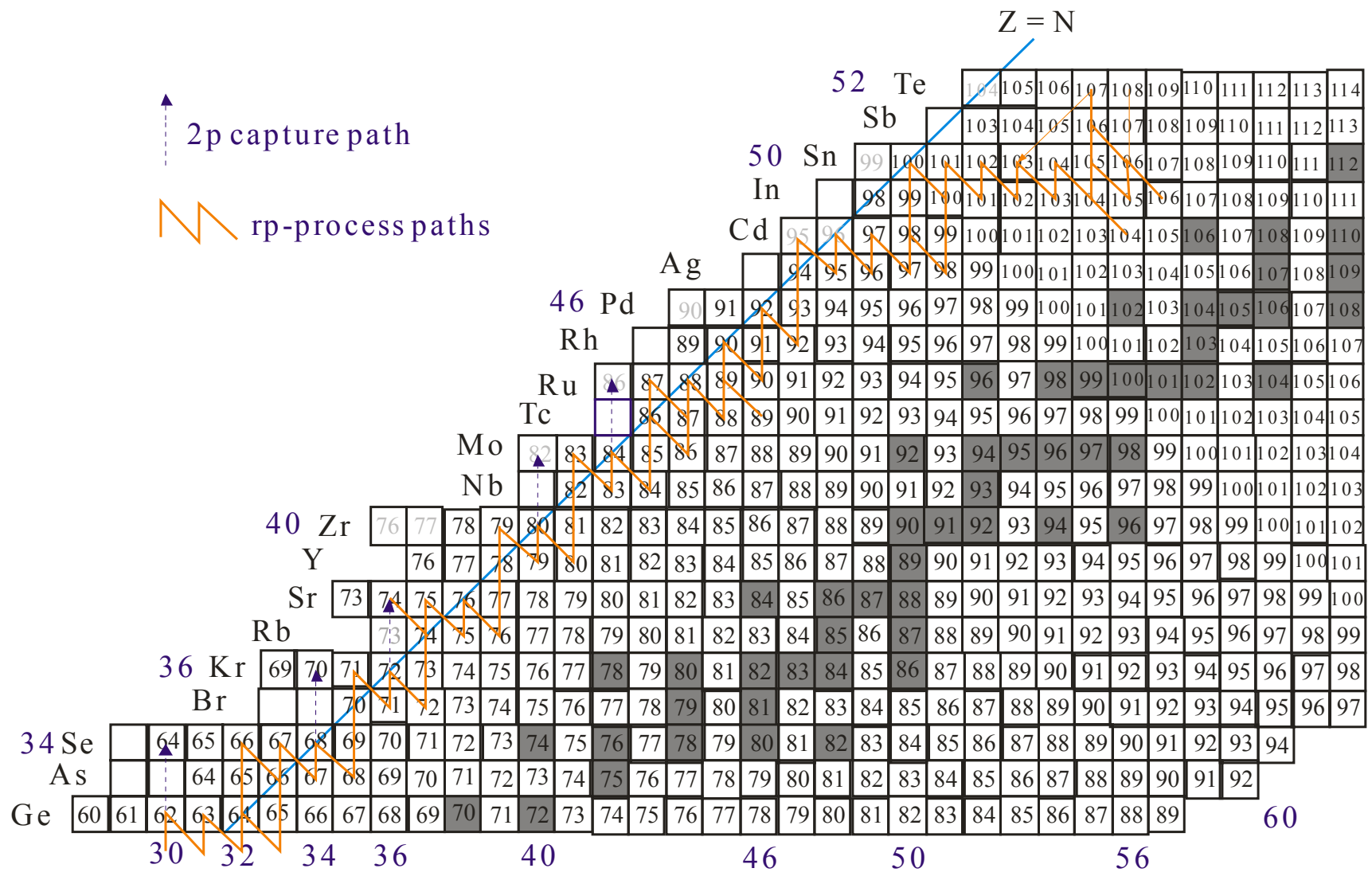

Fig. 4. The chart of nuclides in the mass region of Ge to Te.

Table I. Candidates for the two-proton capture reactions and their intermediate proton-unbound nuclei.

| Elastic resonance scatterings | Intermediate proton-unbound nuclei | Two proton capture reactions | Spin-parity and half-life for the final nucleus |
|---|---|---|---|
| $^{15}O + p$ | $^{16}F$ | $^{15}O\ (2p, \gamma)\ ^{17}Ne$ | $1/2^-$ ;109 ms |
| $^{18}Ne + p$ | $^{19}Na$ | $^{18}Ne\ (2p, \gamma)\ ^{20}Mg$ | $0^+$ ; 90.8 ms |
| $^{20}Mg + p$ | $^{21}Al$ | $^{20}Mg\ (2p, \gamma)\ ^{22}Si$ | $0^+$ ; 9 ms |
| $^{29}S + p$ | $^{30}Cl$ | $^{29}S\ (2p, \gamma)\ ^{31}Ar$ | $5/2^+$ ; 14.4 ms |
| $^{37}Ca + p$ | $^{38}Sc$ | $^{37}Ca\ (2p, \gamma)\ ^{39}Ti$ | $(3/2^+)$ ; 31(+6/-4) ms |
| $^{38}Ca + p$ | $^{39}Sc$ | $^{38}Ca\ (2p, \gamma)\ ^{40}Ti$ | $0^+$ ; 52.4 ms |
| $^{41}Ti + p$ | $^{42}V$ | $^{41}Ti\ (2p, \gamma)\ ^{43}Cr$ | $(3/2^+)$ ; 20.6 ms |
| $^{58}Zn + p$ | $^{59}Ga$ | $^{58}Zn\ (2p, \gamma)\ ^{60}Ge$ | $0^+$; > 110 ns |
| $^{62}Ge + p$ | $^{63}As$ | $^{62}Ge\ (2p, \gamma)\ ^{64}Se$ | $0^+$ ; > 180 ns |
| $^{68}Se + p$ | $^{69}Br$ | $^{68}Se\ (2p, \gamma)\ ^{70}Kr$ | $0^+$ ; > 0.05 s |
| $^{72}Kr + p$ | $^{73}Rb$ | $^{72}Kr\ (2p, \gamma)\ ^{74}Sr$ | $0^+$, > 1.2 μs |

As for nuclear structure studies, we investigate the shape coexistence in the proton-rich Kr nuclei with Z ≤ N. For this purpose, the proposed experiments using $^{70,72}$Kr RIBs employ Coulomb excitation measurements which provide information on the nuclear shape built on the ground state as well as built on the first excited $0^+$ state. It is known that the ground state of $^{72}$Kr would be of oblate in shape while $^{74}$Kr and $^{76}$Kr are known to be prolate in their ground states [25-27]. The proposed experimental methods that would require conversion electron spectroscopy are shown in Fig. 5.

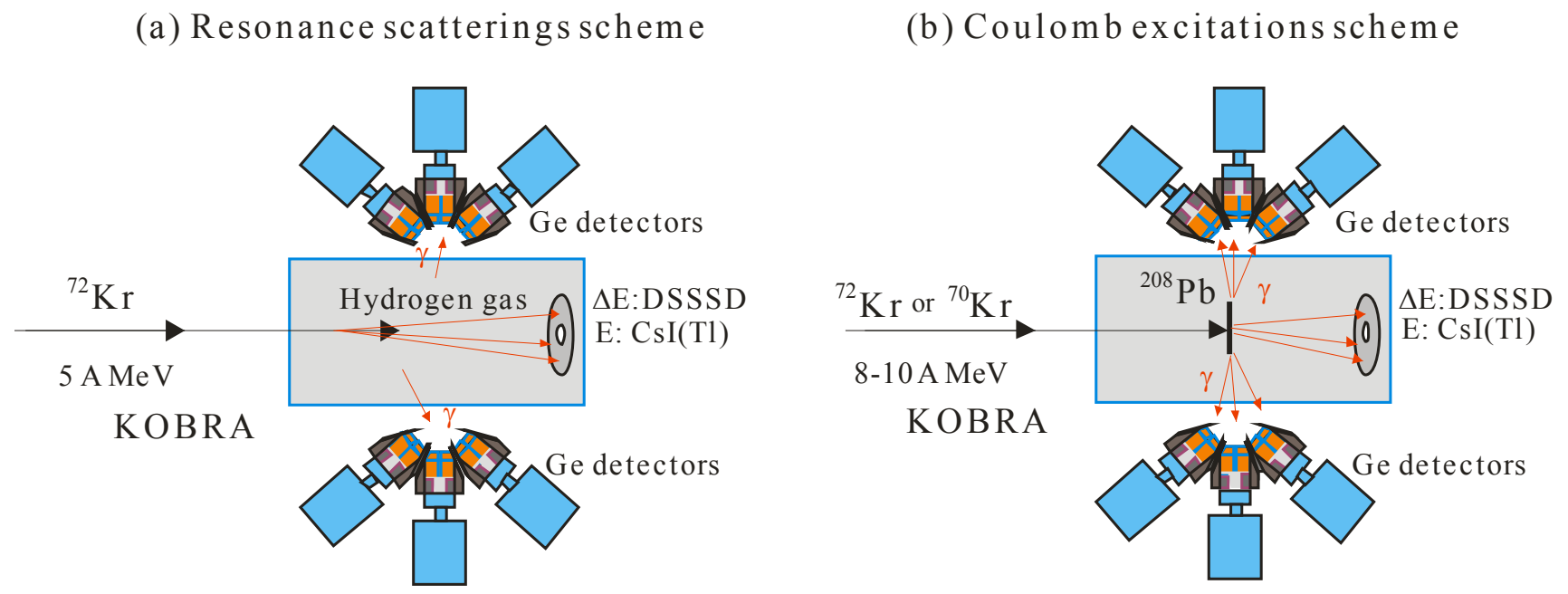

Fig. 5. Experimental set-up schemes for measurements of (a) unbound states in $^{73}$Rb and (b) excited states of $^{72}$Kr and $^{70}$Kr.

In addition, we propose invariant mass measurements for the proton unbound nuclei, such as $^{82}$Mo, $^{86}$Ru, and $^{90}$Pd. They provide information on isospin symmetry, as well as rp-processes beyond the proton drip lines.

**B Study on shell structure evolution along a chain of N = 19, 20, 21 isotones with Z = 22 and 24; $^{41}$Ti, $^{44}$Cr, and $^{45}$Cr.**

This proposal aims at measuring internal structures of the proton-rich $^{41}$Ti, $^{44}$Cr, and $^{45}$Cr nuclei to test the symmetry of mirror states in isobaric A = 41, 44 and 45 by comparing to the structures of $^{41}$K, $^{44}$Ca, and $^{45}$Sc. The nuclei of interest are produced through one or two neutron removal out of secondary beams, knockout reactions, on a thick $^{9}$Be target. The secondary beams are produced by the fragmentations of a 250 MeV per nucleon $^{78}$Kr primary beam on a thin $^{9}$Be production target. The proposed reactions are as follows: $^{9}$Be ($^{42}$Ti, $^{41}$Ti + n) X for $^{41}$Ti, $^{9}$Be ($^{46}$Cr, $^{44}$Cr + 2n) X for $^{44}$Cr, and $^{9}$Be ($^{46}$Cr, $^{45}$Cr + n) X for $^{45}$Cr. The γ-ray spectroscopy employs to measure the internal electromagnetic transitions in the nuclei of interest.

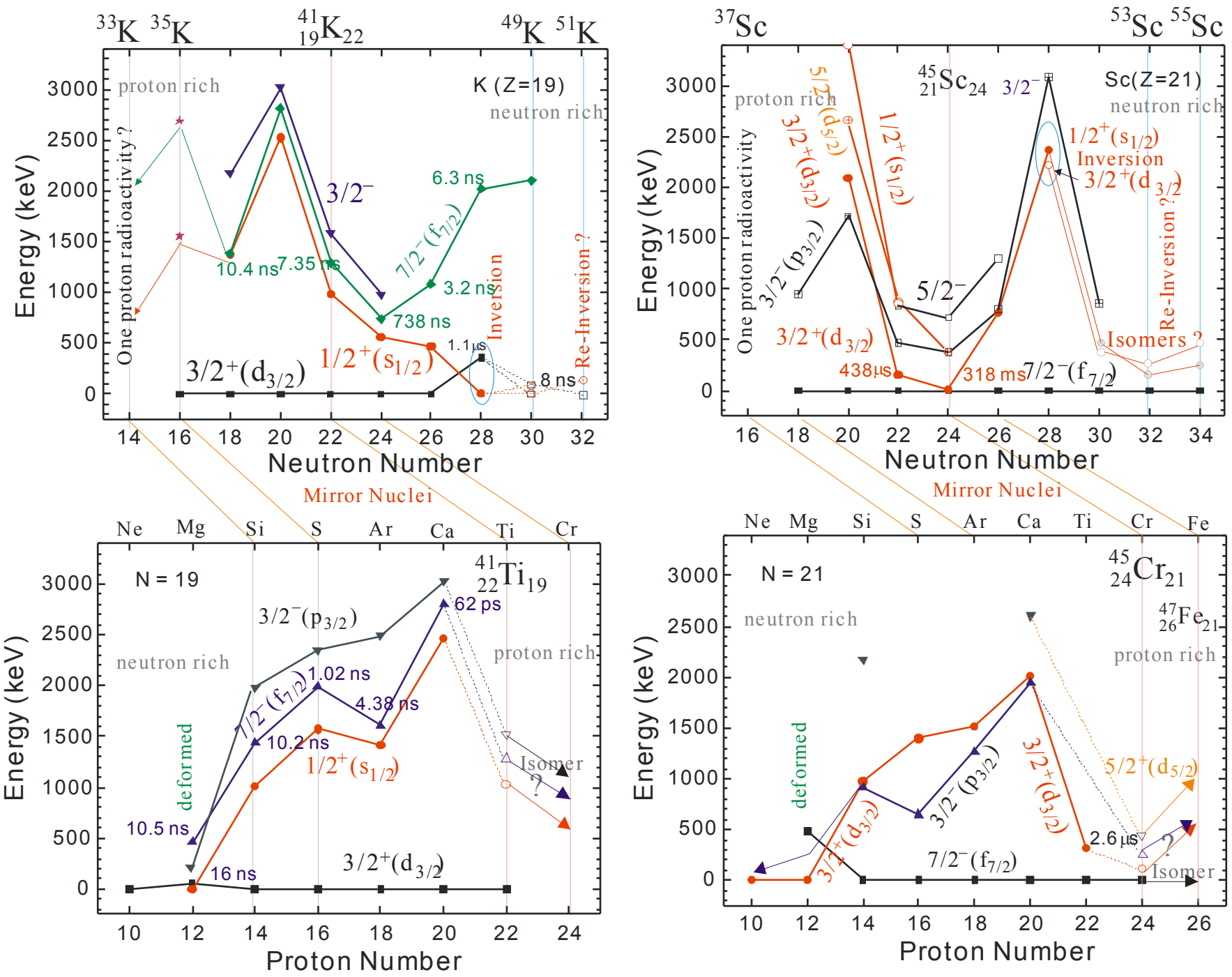

Fig. 6. Systematics of the ground and some low-lying excited states in Z = 19, 21 (upper) as a function of neutron numbers and in N = 19, 21 (lower) as proton numbers [11]. For discussion, the associated level sequences are included as designating their relations by the arrows with the corresponding shell structure evolution.

The goals of the proposal are: (1) Identification of $E(2^+)$ and $E(4^+)$ in even-even $^{44}Cr$ for investigating the magic N = 20 shell evolution like an emergence of collectivity; (2) measurements for the low-lying excited states in odd-Z $^{41}Ti$ and $^{45}Cr$ for testing charge symmetry between mirror nuclei, Z , N = 19 and 21 with A = 41 and 45; and (3) search for isomers in the nuclei of interest to study the single-particle and collective features based on isomerism.

Figure 6 illustrates the energy level systematics for the ground and low-lying single-particle excited states in Z = 19 and 21 nuclei as a function neutron numbers and in N = 19 and 21 as a function of proton numbers. This proposal is focused on the proton rich side. The neutron-rich side will be discussed in plan C. In the region of Z, N = 19 and 21, the $2s_{1/2}$, $1d_{3/2}$, $1f_{7/2}$, and $2p_{3/2}$ orbitlas play a critical role in the strength of the shell gaps as shown in Fig. 6. We find the following distinctive features: First, for Z = 19 isotopes, the $1/2^+$ level due to $2s_{1/2}$ subshell decreases rapidly with increasing neutron numbers from N = 20 and finally becomes the ground state at N = 28. For N = 19 isotones, the subshell also decreases in energy toward ground level with decreasing proton numbers from Z = 20 and finally becomes the ground state at Z = 12. We notice that the nuclei at and below Z = 12 turned out to be deformed in their ground states at magic neutron number 20 [10]. Second, for Z or N = 19, the $7/2^-$ level due to the $1f_{7/2}$ subshell forms isomers with half-lives of a few nanoseconds across the 20 magic gap region as an intruder state. Third, for Z and N = 21, the $3/2^+$ level due to $1d_{3/2}$ subshell changes dramatically with neutron and proton numbers and forms isomers near the ground states at N = 22, 24 and Z = 22. This proposal aims at proving the charge symmetries by observing low-lying states in isobars with Z, N = 19 and 21 toward Z > 20 and N < 18. Besides, we are interested in searching for isomers in $^{41}Ti$ and $^{45}Cr$ as denoted in Fig. 6.

**C Study on shell structure evolution based on the proton single-particle orbital changes in the vicinity of $^{54}$Ca.**

This plan aims at exploring single-particle shell migrations in the vicinity of $^{52}Ca$ and $^{54}Ca$. The experiments are based on the nucleon removal (knockout) reactions using the IFFS. The proposed one-proton removal reactions are as follows: $^{9}Be$ ($^{50}Ca$, $^{49}K$ + p) X, $^{9}Be$ ($^{52}Ca$, $^{51}K$ + p) X, $^{9}Be$ ($^{54}Ti$, $^{53}Sc$ + p) X, and $^{9}Be$ ($^{56}Ti$, $^{55}Sc$ + p) X. This proposal has a goal for studying nuclear structures of the very neutron rich nuclei with Z = 19 and 21 and N ≥ 30; $^{49}K$, $^{51}K$, $^{53}Sc$, and $^{55}Sc$. The measurements involve the thick reaction target $^{9}Be$ and the γ-ray spectroscopy of the projectile-like residual nuclei for the final state resolution. The secondary rare isotopes beams produce from the fragmentation between a $^{86}$Kr primary beam with 250 MeV per nucleon and a $^{9}$Be production target.

As shown in Fig. 6, the $2s_{1/2}$, $1d_{3/2}$ and $1f_{7/2}$ subshells contribute dominantly to shell structure change in proton-single levels. In contrast, neutrons begin to occupy the $2p_{3/2}$, $2p_{1/2}$ and/or $1f_{5/2}$ orbitals cross over the $1f_{7/2}$ orbital, which separates the N = 20 gap and the N = 28 gap. For Z = 19 isotopes, the $1/2^+$ level due to $2s_{1/2}$ subshell decreases remarkably with increasing neutron numbers and finally becomes ground state at and beyond N = 28. For Z = 21 isotopes, the $3/2^+$ level due to $1d_{3/2}$ subshell changes dramatically as it goes down near to the ground state at N = 24. It is interesting to notice that the levels of $3/2^+$ and $1/2^+$ are inverted at N = 28 in Z = 19 ($^{47}$K) and Z = 21 ($^{49}$Sc).

The inversion of the $3/2^+$ and $1/2^+$ levels, however, is uncertain in $^{49}$K and unknown in $^{53}$Sc and $^{55}$Sc above N = 28. It is worthwhile to remind that N = 32 and 34 develop the semi-double shell gaps at Z = 20**.** We address how the distinctive quasiparticle levels, such as $2s_{1/2}$, $1d_{3/2}$, and $1f_{7/2}$ migrate and make an impact on the formation of the semi-double shell closure of $^{54}$Ca. The migration of the proton $2s_{1/2}$, $1d_{3/2}$, $1f_{5/2}$ levels as well as the neutron $2p_{1/2}$, $1f_{5/2}$ levels with respect to neutron numbers may indicate an underlying physics not well accounted for in the present shell model interactions. In this regard, the present proposal is an important step toward extremely neutron-rich region where more exotic phenomena including halos or skins are expected.

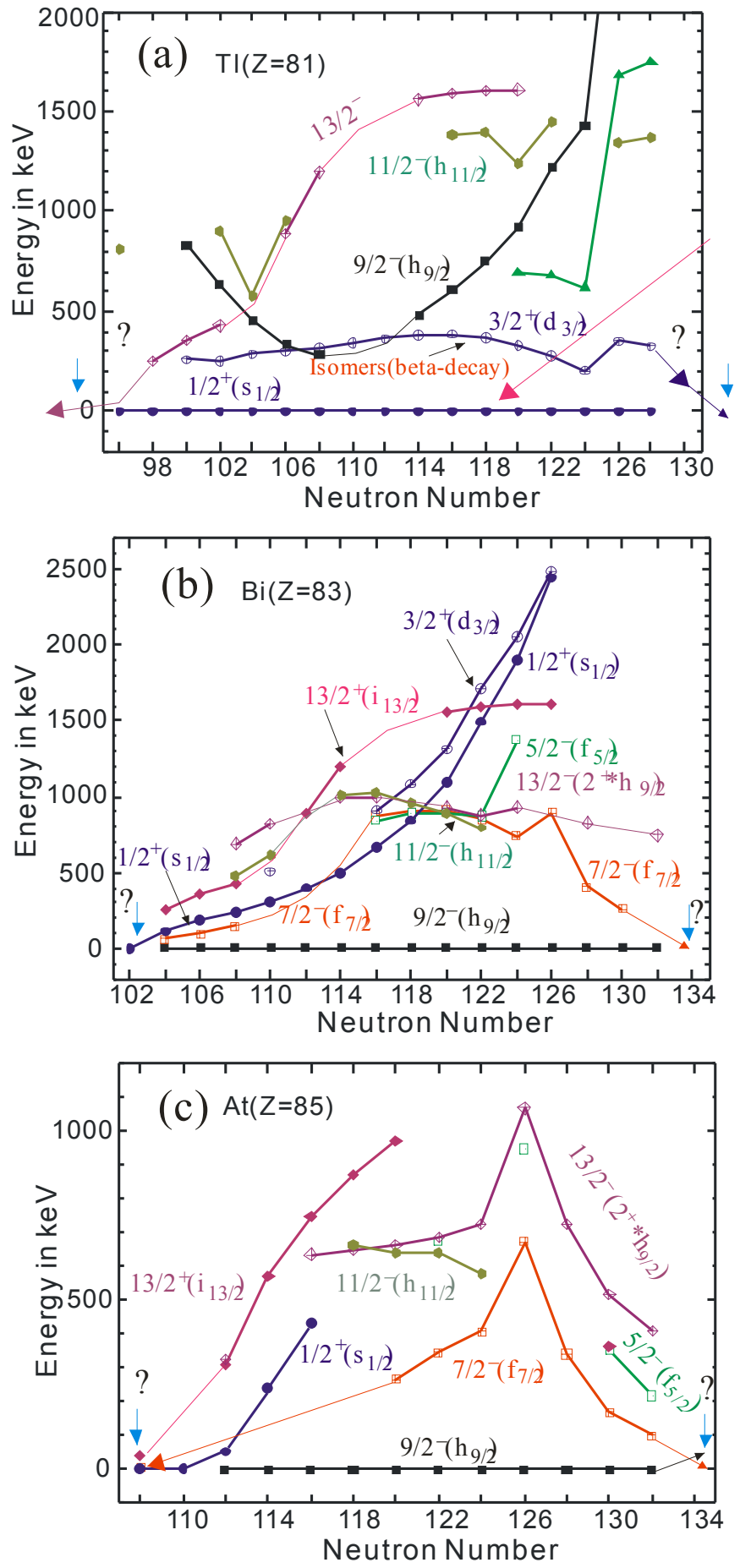

Fig. 7. Systematics of ground and low-lying states in the odd-mass nuclei; (a) Thallium, (b) Bismuth, and (c) Astatine as a function of neutron numbers. The expected level crossings on the basis of the systematic trends are denoted by the arrows. Data are from NNDC [12].

### D Shell structure evolution and shape transitions in the vicinity of proton magic number, Z = 82.

This region is a section of the nuclear chart that is not easy to access under current experimental condition. In particular, both the north-eastern area and the south area of the Z = 82 and N =126 point remain mostly unexplored. As a starting point to study shell structure evolution based on the single particle configurations over Z = 82 and N = 126, we focus on investigating the level structures of Tl (Z = 81) and Bi (Z = 83) since the intruder states and the associated isomers in odd Z nuclei with ±1 outside doubly magic core provide critical information on the shell structure evolution. Figure 7 demonstrates the single-particle energy level systematics based on the low-lying excited states for odd-Z nuclei lying close to $^{208}$Pb. By considering the currently known nuclei, we focus on the following nuclei: $^{175}Tl_{94}$, $^{177}Tl_{96}$, $^{179}Tl_{98}$, $^{211}Tl_{130}$, $^{213}Tl_{132}$, $^{183}Bi_{100}$, $^{185}Bi_{102}$, $^{217}Bi_{134}$, $^{219}Bi_{136}$, $^{191}At_{106}$, $^{193}At_{108}$, $^{219}At_{134}$, and $^{221}At_{136}$. An interesting feature is that the levels at $J^{\pi} = 1/2^{+}$, $7/2^{-}$, as well as $13/2^{+}$ decrease down to the ground level as the neutron numbers decrease in Bi and At nuclei. It is also interesting to know whether the $7/2^{-}$ level would reach the ground state beyond N = 132 in the said nuclei. The shell migration of these orbitals is closely connected to concepts of the triplet shape coexistence in the region of neutron deficient Pt, Os, Hg, and Pb.

The nuclei of interest can be investigated by the multi-transfer reactions at E ~ 10 MeV per nucleon. For example the reaction of $^{136}$Xe + $^{208}$Pb at $E_{cm}$ = 526 MeV is estimated to be 0.2 μb for the production of $^{219}$At and 3 μb for the $^{217}$Bi, respectively. However, the role of transfer channels is less clear due to the difficulty of making qualitative calculations for both the multi-nucleon transfer and the sub-barrier fusion simultaneously. The very neutron-rich RIBs, for example $^{144}$Xe, with the high intensity and high quality with a few MeV per nucleon are essential for producing exotic nuclei far from the Z = 82 and N = 126 point, such as $^{196}$Yb and $^{220}$Pt. It is noteworthy to mention that the measurements of half-lives of the nuclei with Z < 82 and N = 126 provide decisive information on the r-process paths along waiting points with N = 126, while the β-delayed neutron emission schemes in these neutron rich-nuclei play an important role in determining abundances of Au and Pt elements.

## 6 Summary and outlook

We have introduced and discussed the planned experiments using RIBs for the Korean Rare Isotope Beams Accelerator facility, RAON to be built in a decade. This facility is one of the world-class multi-beam facilities capable of producing and accelerating beams of wide range mass of nuclides with energies of a few to hundreds MeV per nucleon. The low energy RIBs at $E_{lab}$ = 5 to 20 MeV per nucleon are used for the study of nuclear structures and nuclear astrophysics toward and beyond the drip lines while the higher energy RIBs produced by the in-flight fragmentation with the reaccelerated ions from the ISOL enable us to explore the neutron drip lines in intermediate mass regions. Beam specifications for the KOBRA and ones for the IFFS spectrometers are complementary, which allow scientists to extent the investigations toward the very neutron-rich and proton-rich nuclei.

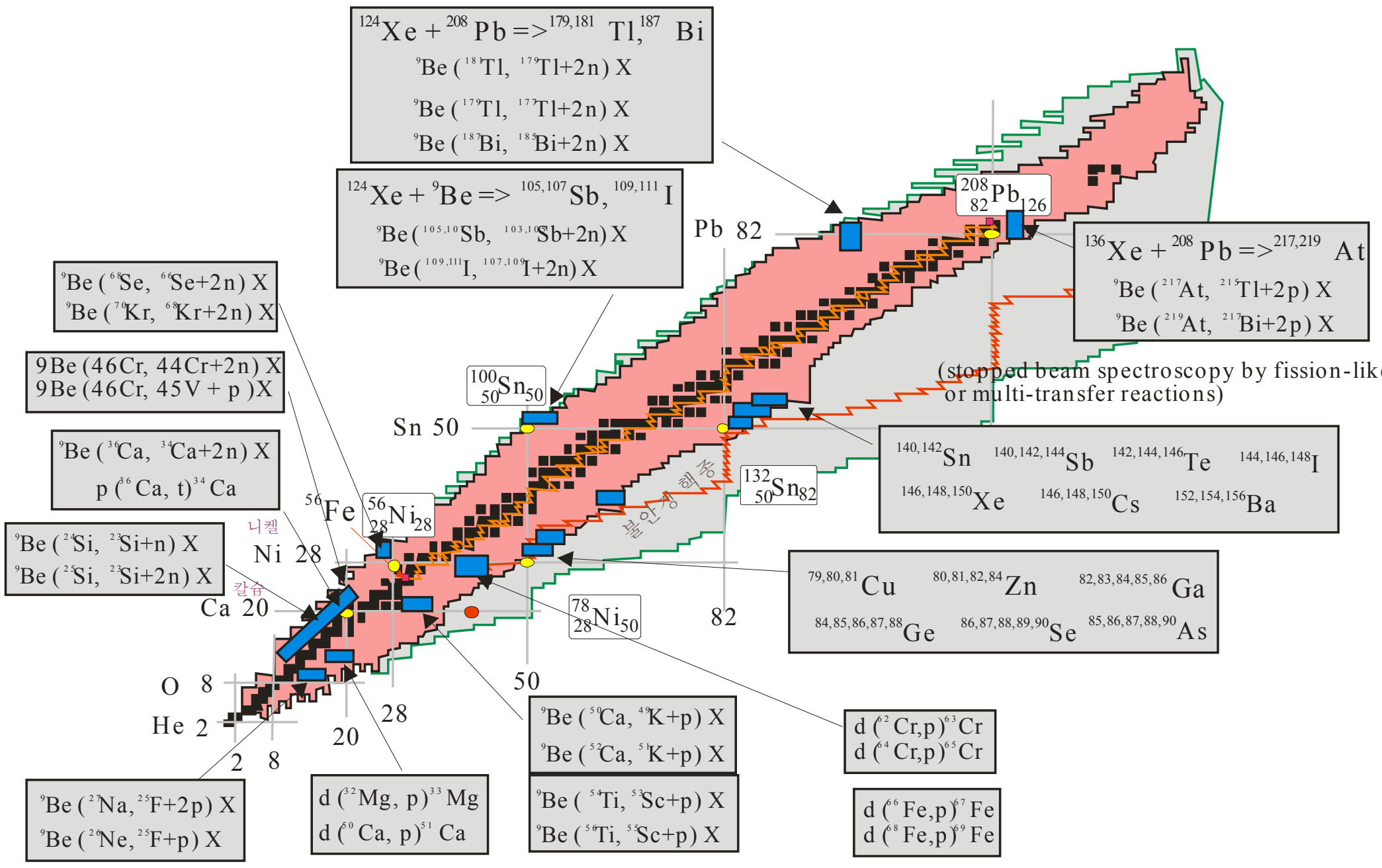

Fig. 8. Summary of the proposed experiments aiming at the rare isotope beams accelerator facility [11], RAON.

The experiments are planned to investigate the internal structures of exotic nuclei toward and beyond the nucleon drip lines as addressing the following questions: how the shell structure evolves in areas of extreme proton to neutron imbalance; whether the isospin symmetry maintains in isobaric mirror nuclei at and beyond the drip lines, how two-proton or two-neutron radioactivity affects abundances of the elements, what the role of the continuum states including resonant states above particle-decay threshold in exotic nuclei is in astrophysical nuclear reaction processes, and how the nuclear reaction rates triggered by unbound proton-rich nuclei make an effect on rapid proton capture processes in a very hot stellar plasma.

To summarize, we show a set of the planned experiments with the associated nuclei to be investigated over the nuclear mass chart in Fig. 8. We hope that the proposed experiments will play an important role in developing a nuclear physics program for the rare isotope beams accelerator facilities, as well as in promoting the creation of next generation of nuclear scientists in the world.

* This is not an official abbreviation. The Rare Isotope Beams Accelerator Facility to be built in Korea has been named officially, regardless of an abbreviation for the facility, as 'RAON' which is said to mean 'joyful, delightful' in ***old*** Korean language.

**References**

[1] http://www.ganil-spiral2.eu/.
[2] http://www.rarf.riken.go.jp/.
[3] http://nscl.msu.edu/.
[4] http://frib.msu.edu/.
[5] http://www.gsi.de/.
[6] http://www.triumf.ca/.
[7] http://isolde.web.cern.ch/ISOLDE/.
[8] Heyde K, Isacker P van, Waroquier M, Wood J L, and Meyer R A, Phys. Rep. 1983, 102, 291.
[9] Wood J L, Heyde, Nazarewicz K.W, Huyse M, and Duppen P van, Phys. Rep. 1992, 215, 101.
[10] Heyde K and Wood J L, Rev. Mod. Phys. 2011, 83, 1467.
[11] Moon Chang-Bum, AIP Adv. 2014, **4**: 041001.
[12] National Nuclear Data Center, Brookhaven National Laboratory, http:// www.nndc.bnl.gov/.
[13] Riisager K, Butler P, Huyse M, and Krucken R, *HIE-ISOLDE: the scientific opportunities*, CERN-2007-008.
[14] Arnett David, *Supernovae and Nucleosynthesis*, Princeton University Press: Princeton, 1996.
[15] Heyde H, *Basic Ideas and Concepts in Nuclear Physics*, IOP Publishing, Bristol and Philadelphia, 1999.
[16] Pfutzner M, Karny M, Grigorenko L V, and Riisager K, Rev. Mod. Phys. 2012, **84**: 567.
[17] Kratz Karl-Ludwig, AIP Conf. Proc. 2015, **1645** :109.
[18] Bertulani Carlos, *9th Symposium on Nuclei in the Cosmos* (NIC IX), Proceedings of Science(PoS), 2006: 040.
[19] Orr N, *Physics with Reaccelerated ISOL Beams*, J. Phys. G: Nucl. Part. Phys. 2011, **38**: 020301.
[20] Schatz H, Aprahamian A, Görres J, Wiescher M, Rauscher T, Rembges J F, Thielemann F -K, Pfeiffer B, Möller P, Kratz K -L, H. Herndl, B. A. Brown, and H. Rebel, Phys. Rep. 1998, **294**: 167.
[21] Ravn H L and Allardyce B W, *Treatise Heavy Ion Science*, Vol. 8, Ed. D. A. Bromley, Plenum Press, New York and London, 1989: 363-439.
[22] Tanihata Isao, *Treatise Heavy Ion Science*, Vol. 8, Ed. D. A. Bromley (Plenum Press, New York and London, 1989: 443-514.
[23] Progress report for the RAON recoil spectrometer KOBRA, **2015**.
[24] Tshoo K, Chae H, Park J, Moon J Y, Kwon Y K, Souliotis G A, Hashimoto T, Akers C, Berg G P A, Choi S, Jeong S C, Kato S, Kim Y K, Kubono S, Lee K B, and Moon C-B, Proceedings of EMIS2015 (17th International Conference on Electromagnetic Isotope Separators and Related Topics), to be published in Nuclear Instruments and Methods.
[25] Bouchez E et al., Phys. Rev. Lett. 2003, **90** : 082502.
[26] Poirier E et al., Phys. Rev. C 2004, **69** : 034307.
[27] Clément E et al., Phys. Rev. C 2007, **75** : 054313.